\documentclass[showpacs]{revtex4}

\usepackage{graphics}
\usepackage{amssymb}
\usepackage{amsmath}

\newcommand{\p}{\partial}

\begin{document}
\title{Integration of the geodesic equations via Noether Symmetries}

\author{Ugur Camci}

\email{ucamci@rwu.edu,ugurcamci@gmail.com}

\bigskip

\affiliation{Department of Chemistry and Physics, Roger Williams University, Bristol, Rhode Island 02809, USA}

\bigskip

\date{\today}


\begin{abstract}
Through this article, I will overview the use of Noether symmetry approach in discussing the integration of geodesic equations for the geodesic Lagrangians of spacetimes. I will also give some examples to reveal the efficiency of Noether symmetry approach by finding the first integrals related for the geodesic Lagrangians of the G\"{o}del-type, Schwarzschild, Reissner-Nordstr\"{o}m and Kerr spacetimes. After obtaining the approximate Noether symmetries of the Schwarzschild, Reissner-Nordstr\"{o}m and Kerr spacetimes, the first integrals associated with each of approximate Noether symmetries have been integrated to find a general solution of geodesic equations in terms of the arc length $s$.
\end{abstract}



\maketitle

\section{Introduction}\label{intr:sec1}

The differential equations can be deduced from a Lagrangian function through a variational technique.  Noether symmetries \cite{noether} which are the special classes of Lie symmetries are intimately connected with conservation laws, {\it or} the first integrals in the case of ordinary differential equations (ODEs) derived from the corresponding Lagrangian.
The equations of geodesic motion are expressed in terms of  the configuration space variables, i.e., the metric coefficients. Therefore, the configuration space of our model is a $4-$dimensional Riemannian manifold with coordinates $x^i \,(i=0,1,2,3)$, in which we  construct a point-like geodesic Lagrangian to produce the geodesic equations of motion.
The geodesic equations are a system of second order ODEs and can be derived from a Lagrangian function $\mathcal{L} (\tau, x^i, \dot{x}^i)$ of the system related to the geodesic motion. Here the dot represents the derivative with respect to an affine parameter $\tau$, and this is the arc length $s$ in most of the spacetimes. Note that $Q = \{ x^i, i=0,1,2,3 \}$ is the configuration space from which it is possible to derive the corresponding tangent space $ TQ= \{ x^i, \dot{x}^i \}$ where the Lagrangian $\mathcal{L}$ is defined. Taking the variation of the geodesic Lagrangian
\begin{equation}
\mathcal{L} (s, x^k, \dot{x}^k) = \frac{1}{2} g_{ij} (x^k) \dot{x}^i \dot{x}^j - V(s,x^k) \, , \label{canonic-lagr}
\end{equation}
with respect to the coordinates $x^i$, it follows the geodesic equations of motion,
\begin{equation}
\ddot{x}^i +  \Gamma^i_{j k} \dot{x}^j \dot{x}^k  = F^i (s, x^{\ell} ), \label{ode}
\end{equation}
where $\Gamma^i_{j k} ( x^{\ell} )$ are the connection coefficients of the metric and $ F^i = g^{i j} V_{,j}$ is the conservative force field.
The \emph{energy functional} associated with $\mathcal{L}$ is
\begin{equation}
E_{\mathcal{L}} = \dot{x}^j \frac{\partial \mathcal{L}}{\partial \dot{x}^j} - \mathcal{L} = \frac{1}{2} g_{i j} \dot{x}^i \dot{x}^j + V(s, x^k) \, , \label{energy}
\end{equation}
which is the Hamiltonian of the system.

Noether symmetries are associated with differential equations possessing a Lagrangian, and they describe physical features of differential equations in terms of conservation laws admitted by them \cite{ibragimov}. Thus one can use the geodesic Lagrangian associated with the geodesic motion for spacetimes to integrate the geodesic equations of motion if it is possible. It is capable to construct analytical solutions of geodesic equations by reducing their complexity, using not only the Noether symmetry but also the approximate Noether symmetry approach. In order to find out analytical solutions of geodesic equations for the considered geodesic Lagrangian, one can use the obtained first integrals of motion which are Noether constants. Recently the Noether symmetries of geodesic Lagrangian for some spacetimes have been calculated, and classified according to their Noether symmetry generators \cite{feroze1}-\cite{hussain2020}.
In this study we consider Noether symmetries (the approximate Noether symmetries) of the geodesic Lagrangian (the perturbed geodesic Lagrangian) for the line elements of some known spacetimes, rather than those of the geodesic equations.

We design this study as follows. In the following section, we aim to give some examples of Noether symmetries for the geodesic Lagrangian $\mathcal{L}$ of some well-known spacetimes.
In Section \ref{sec3}, we present a detailed analysis of the approximate Noether symmetries of Schwarzschild, Reissner-Nordstr\"{o}m  and Kerr spacetimes. Finally, we conclude with a brief summary and discussions in Section \ref{conc:sec4}.

\section{Noether Symmetries for the geodesic Lagrangians}\label{sec2}

The Noether symmetry (NS) generator for the geodesic Lagrangian associated with the system of ODEs in \eqref{ode} is
\begin{eqnarray}
& & {\bf X} = \xi (s, x^k) \frac{\partial}{\partial \tau} + \eta^i ( s, x^k) \frac{\partial}{\partial x^i} \, ,  \label{ns-generator}
\end{eqnarray}
if there exists a function $f(s, x^k)$ which is sometimes called a gauge function, and
the NS condition
\begin{equation}
{\bf X}^{[1]} \mathcal{L} + \mathcal{L} ( D_{s} \xi )= D_{s} f  \, ,  \label{ngs-cond}
\end{equation}
is satisfied, where $D_{s} = \partial / \partial s + \dot{x}^j \partial / \partial x^j$ is the total derivative operator and ${\bf X}^{[1]}$ is the first prolongation of NS generator ${\bf X}$, i.e.
\begin{equation}
{\bf X}^{[1]} = {\bf X} + \dot{\eta}^j (s, x^{\ell}, \dot{x}^{\ell}) \frac{\partial}{\partial \dot{x}^j}   \,  ,   \label{first-pro}
\end{equation}
with $\dot{\eta}^j (s, x^{\ell}, \dot{x}^{\ell}) = D_{s} \eta^j - \dot{x}^j D_{s} \xi$.
It has to be noted here that the NS condition \eqref{ngs-cond} takes the alternative form
\begin{eqnarray}
& & \xi_{,i} = 0, \qquad g_{i j} \eta^j_{,s} = f_{,i} \, , \qquad  \pounds_{\bf \eta} g_{ij} = \xi_{,s} g_{ij} \, ,  \qquad  \pounds_{\bf \eta} V = - \xi_{,s} V - f_{,s} \, , \label{neq-1234}
\end{eqnarray}
where $\pounds_{\bf \eta}$ is the Lie derivative operator along ${\bf \eta}$ ,and
the set of all NSs form a finite dimensional Lie algebra denoted by $\mathcal{N}$.

For every NS, there is a {\it conserved quantity} (or {\it a first integral}) of the system of equations \eqref{ode} given by
\begin{equation}
I = - \xi E_{\mathcal{L}}+ g_{i j} \eta^i  \dot{x}^j - f \, ,  \label{first-int}
\end{equation}
where the {\it energy functional} $E_{\mathcal{L}}$ of the geodesic Lagrangian is the same as in Eq. \eqref{energy}.

Now let us recall the spacetime symmetries. A vector field ${\bf K}$, satisfying the equation \cite{katzin}
\begin{equation}
\pounds_{\bf K} g_{ij} = 2 \psi g_{ij} \, , \label{killing}
\end{equation}
is called a conformal Killing vector (CKV), where $\pounds_{\bf K}$ is the Lie derivative operator along the vector field ${\bf K}$, and $\psi = \psi (x^i)$ is a conformal factor. For $\psi_{;ij} \neq 0$, the CKV field ${\bf K}$ is said to be {\it proper}, otherwise it is a special conformal Killing vector (SCKV) field ($\psi_{;ij} = 0$). The vector field ${\bf K}$ is a homothetic vector (HV) for $\psi_{,i} = 0$, and it is an isometry or a Killing vector (KV) field for $\psi = 0$. The set of all CKV (respectively SCKV, HKV and KV) form a finite-dimensional Lie algebra denoted by $\mathcal{C}$ (respectively $\mathcal{S}, \mathcal{H}$ and $\mathcal{G}$).

\subsection{Noether Symmetries of the G\"{o}del-type Spacetimes }

In cylindrical coordinates $x^i = (t,r,\phi,z), \, i= 0,1,2,3$, the line element for the G\"{o}del-type spacetimes can be written as
\begin{equation}
ds^2 = \left[ dt + H(r) d \phi \right]^2 - dr^2 - D(r)^2 f\phi^2 - dz^2. \label{godel}
\end{equation}
The necessary and sufficient conditions for a G\"{o}del-type manifold to be spacetime homogeneous (STH) are given by
\begin{equation}
\frac{D''}{D} = {\rm const.} \equiv m^2, \qquad \frac{H'}{D} = {\rm const.} \equiv - 2 \omega \, , \label{godel-eqs}
\end{equation}
where prime ($ ' $)  denotes derivative with respect to the radial coordinate $r$.
All STH Riemannian manifolds endowed with the G\"{o}del-type spacetime \eqref{godel} are obtained by solving equations in \eqref{godel-eqs} as follows: \\
{\bf Class I:} $m^2 > 0, \omega \neq 0$.
\begin{equation}
H(r) = \frac{2 \omega}{m^2} \left[ 1 - \cosh(m r) \right], \qquad D(r) = \frac{1}{m} \sinh(m r) .
\end{equation}
{\bf Class II :} $ m^2 = 0, \omega \neq 0$.
\begin{equation}
H(r) = -\omega r^2  \, , \qquad D(r) = r  \, .
\end{equation}
{\bf Class III :} $ m^2 \equiv - \mu^2 < 0, \omega \neq 0$.
\begin{equation}
H(r) = \frac{2 \omega}{\mu^2} \left[ \cos(\mu r) -1 \right], \qquad D(r) = \frac{1}{\mu} \sin(\mu r) .
\end{equation}
{\bf Class IV :} $ m^2 \neq < 0, \omega = 0$. This class yields a degenerate G\"{o}del-type manifolds, since the cross term related to the rotation $\omega$ in the line element vanishes.
One can make $H = 0$ by a trivial coordinate transformation with $D(r)$ given as in Classes I and III depending on whether $m^2 > 0$ or $m^2 \equiv -\mu^2 < 0$.

Using the G\"{o}del-type spacetime \eqref{godel}, the geodesic Lagrangian takes such a form
\begin{equation}
\mathcal{L} = \frac{1}{2} \left[ \dot{t}^2 - \dot{r}^2 - \dot{z}^2 + (H(r)^2 - D(r)^2) \dot{\phi}^2 \right] + H(r) \dot{t} \dot{\phi} - V(t,r,\phi,z) \, .
\end{equation}
Then it follows for this Lagrangian that the energy functional is
\begin{equation}
E_{\mathcal{L}} = \frac{1}{2} \left[ \dot{t}^2 - \dot{r}^2 - \dot{z}^2 + (H(r)^2 - D(r)^2) \dot{\phi}^2 \right] + H(r) \dot{t} \dot{\phi} + V(t,r,\phi,z) \, ,
\end{equation}
and the conserved quantity associated with NS generator ${\bf X}$ is
\begin{equation}
I = - \xi E_{\mathcal{L}} + \left( \eta^0 + H \eta^2 \right) \dot{t} - \eta^1 \dot{r} + \left[ H \eta^0 + ( H^2 - D^2) \eta^2 \right] \dot{\phi} - \eta^3 \dot{z}  - f(s,t,r,\phi,z) \, . \label{fint-godel}
\end{equation}

The complete NS analysis of G\"{o}del-type spacetimes for classes I, II, III and IV has been given by Camci\cite{ug2014,ug2015}. Let us briefly summarize the results.
The geodesic Lagrangian $\mathcal{L}$ of G\"{o}del-type spacetimes for classes I, II, III and IV yields  {\it 7}  NS generators. Thus, the G\"{o}del-type spacetimes corresponding to those classes admit the algebra $\mathcal{N}_7 \supset \mathcal{G}_5$.
In special  class I  (where $ m^2 = 4 w^2$) and class IV,  it is found {\it 9} NS generators. The NS algebra admitted by the special class I is $\mathcal{N}_9 \supset \mathcal{G}_7$ while the G\"{o}del-type spacetime in class IV admits the algebra $\mathcal{N}_9 \supset \mathcal{G}_6$. The first integrals have been obtained by using the geodesic Lagrangians for the G\"{o}del-type spacetimes of each classes I, II, III and IV, due to the existence of NS vector fields including the KVs. Using the first integrals obtained in all classes of G\"{o}del-type spacetimes, the analytical solutions of geodesic equations have been derived. This result represents the usefulness of the NSs.

As an example, we give only the obtained NSs and associated first integrals for class I as
follows (See Ref. \cite{ug2014} for the details of calculation). For class I there are {\it seven} NSs such that  ${\bf X}_1,..., {\bf X}_5$ are KVs,
\begin{eqnarray}
& &  {\bf X}_1 = \partial_t, \quad {\bf X}_2 = \partial_z, \quad
{\bf X}_3 = \frac{2 \omega}{m} \partial_t - m \partial_{\phi},
\nonumber \\ & &  {\bf X}_4 = - \frac{H}{D} sin\phi \partial_t + cos\phi
\partial_r - \frac{D'}{D} sin\phi \partial_{\phi}, \label{KV-I1} \\
& &  {\bf X}_5 = - \frac{H}{D} cos\phi \partial_t - sin\phi
\partial_r - \frac{D'}{D} cos\phi \partial_{\phi}, \nonumber
\end{eqnarray}
and ${\bf Y}_1,{\bf Y}_2$  are {\it two} non-Killing NSs,
\begin{eqnarray}
& &  {\bf Y}_1 = \partial_{s} \, , \qquad {\bf Y}_2 = s \partial_z \quad {\rm with } \,\, f = -z \, .  \label{Y12}
\end{eqnarray}
Then the first integrals associated with ${\bf X_1},...,{\bf X_5}, {\bf Y_1}$ and ${\bf Y_2}$ are found by the relation \eqref{fint-godel}  as
\begin{eqnarray} \label{frstI-123}
& &  I_1 = \dot{t} + H \dot{\phi}, \quad I_2 = -\dot{z}, \quad I_3 = \frac{2w}{m} I_1 -m \left[ H \dot{t} + (H^2-D^2)\dot{\phi} \right],
\end{eqnarray}
\begin{eqnarray} \label{frstI-4}
& &  I_4 = -\frac{\sin\phi}{D} \left\{ H(1 + D') \dot{t}  + \left[ H^2 + (H^2 -D^2) D'\right] \dot{\phi} \right\} - \cos\phi \, \dot{r},
\end{eqnarray}
\begin{eqnarray} \label{frstI-5}
& &   I_5 =  -\frac{\cos\phi}{D} \left\{ H(1 + D') \dot{t}  + \left[ H^2 + (H^2 -D^2) D'\right] \dot{\phi} \right\} + \sin\phi \, \dot{r},
\end{eqnarray}
\begin{eqnarray} \label{frstI-67}
& &   I_6 =  -E_{\mathcal{L}}, \qquad I_7 = - s \dot{z} + z,
\end{eqnarray}
where the $E_{\mathcal{L}}$ is the Hamiltonian of the dynamical system and yields
\begin{eqnarray} \label{EL-cI}
& &  E_{\mathcal{L}} = \frac{1}{2} \left\{ I_1^2 - I_2^2 - \frac{1}{D^2} \left[ H I_1 - \left( \frac{2w}{m^2}I_1 - \frac{I_3}{m} \right) \right]^2 - \dot{r}^2  \right\}.
\end{eqnarray}
After integrating the above first integrals \eqref{frstI-123}-\eqref{frstI-67}, the general solution can be obtained as follows:
\begin{eqnarray}\label{soln-deq1-cI}
& & z(s) = - p_z s + I_7 \, , \nonumber \\ & & u (s) = \frac{1}{2 \eta} \left[ 1-  \beta^2 + 2 w \gamma + \sqrt{(1- \beta^2 + 2 w \gamma)^2-\eta m^2 \gamma^2} \sin ( m p_t \sqrt{\eta} (s- s_0) ) \right], \nonumber \\ & & t(s) =  \frac{2 w (\gamma + 4 w / m^2)}{m \sqrt{\eta} \sqrt{(1 + p)^2 -q^2}} \arctan \left[ \frac{(1+ p) \tan(m p_t \sqrt{\eta} (s- s_0) /2 ) + q }{\sqrt{ (1+ p)^2 -q^2}} \right] + p_t \left( 1- \frac{4 w^2}{m^2} \right) s + t_0,  \nonumber \\ & &  \phi (s) = \frac{ m (\gamma + 4 w / m^2)}{2 \sqrt{\eta} \sqrt{(1+p)^2 -q^2}} \arctan \left[ \frac{(1+ p) \tan(m p_t \sqrt{\eta} (s- s_0) /2 ) + q }{\sqrt{ (1+ p)^2 -q^2}} \right] \nonumber \\& &  \qquad \qquad - \arctan \left[ \frac{2 \sqrt{\eta}}{m \gamma} \left\{  p \tan (m p_t \sqrt{\eta} (s - s_0)/2 ) + q \right\} \right] + \phi_0, \nonumber
\end{eqnarray}
where $u$ is a new variable defined by $u = m^2 H/4 w$ which is equivalent to $\sinh^2 (m r/2)$ for the {class I}, $\eta \neq 0$, $(1- \beta^2 + 2 w \gamma)^2-\eta m^2 \gamma^2 \geq 0$, $(1+ p)^2 > q^2, \, t_0 = t(0)$ and $\phi_0 = \phi (0)$ .
Here the constants of motion $p_t = I_1,$ $ p_{\phi} = 2 \omega I_1 / m^2$ and $p_z = I_2$ represent the {\it conservation of energy}, {\it angular momentum} and $z$ {\it component of momentum}, respectively, and we have introduced the parameters $p$ and $q$ such as
\begin{eqnarray}
p:= \frac{1- \beta^2 + 2 w \gamma}{2 \eta}, \qquad q := \sqrt{p^2- \frac{m^2 \gamma^2}{4 \eta}}, \nonumber
\end{eqnarray}
where $ p^2 \geq m^2 \gamma^2 / 4 \eta$.

\subsection{Noether Symmetries of the Spherically Symmetric Spacetimes }

The field of a spherically symmetric gravitational source at rest at the origin is given by the Schwarzschild line element
\begin{eqnarray}
& &  ds^2 = \left( 1 - \frac{2 G M}{c^2 r} \right) c^2 dt^2 - \frac{ dr^2}{ \left( 1 - \frac{2 G M}{c^2 r} \right)} - r^2 d\Omega^2, \label{sch-metric}
\end{eqnarray}
where $d\Omega^2 \equiv d\theta^2 + \sin^2 \theta d\phi^2 $, $G$ is Newton's gravitational constant, $M$ is the mass of the point gravitational source and $c$ is the speed of light in vacuum.

The geodesic Lagrangian of the Schwarzschild metric
\begin{eqnarray}
& &  \mathcal{L} = \frac{1}{2} \left[ \left( 1 - \frac{2 G M}{c^2 r} \right) c^2 \dot{t}^2 - \frac{\dot{r}^2}{ \left( 1 - \frac{2 G M}{c^2 r} \right)} - r^2 (\dot{\theta}^2 + \sin^2 \theta \, \dot{\phi}^2 ) \right] - V(t,r,\theta,\phi)  \, ,  \nonumber
\end{eqnarray}
has {\it five} NSs \cite{kmq2008,tsamparlis1} for constant potential, which are {\it four KVs} corresponding the conservation of energy and angular momentum only, i.e.,
\begin{eqnarray}
& & {\bf K}_0 = \p_t, \qquad {\bf K}_1 = \cos \phi \p_{\theta} - \cot \theta \sin \phi \p_{\phi}, \qquad  {\bf K}_2 = \sin \phi \p_{\theta} + \cot\theta \cos\phi \p_{\phi}, \qquad  {\bf K}_3 = \p_{\phi}, \label{kv-0123}
\end{eqnarray}
and the translation of the arc length $s$, i.e., ${\bf Y}_0 = \p_s$.
Note that conservations of linear momentum and the spin angular momentum are {\it lost}.

The Reissner-Nordstr\"{o}m (RN) metric is a static, spherically symmetric and asymptotically flat spacetime
\begin{eqnarray}
& &  ds^2 = \left( 1- \frac{2 G M}{c^2 r} + \frac{G Q^2}{c^4 r^2} \right) c^2 dt^2 - \frac{dr^2}{ 1- \frac{2 G M}{c^2 r} + \frac{G Q^2}{c^4 r^2}  } - r^2 d\Omega^2 , \label{rn-metric}
\end{eqnarray}
where $Q$ is the electric charge of the point gravitational source.

The geodesic Lagrangian of the Reissner-Nordstr\"{o}m metric
\begin{eqnarray}
& &  \mathcal{L} = \frac{1}{2} \left[ \left( 1 - \frac{2 G M}{c^2 r} + \frac{G Q^2}{c^4 r^2} \right) c^2 \dot{t}^2 - \frac{\dot{r}^2}{ \left( 1 - \frac{2 G M}{c^2 r} +  \frac{G Q^2}{c^4 r^2} \right)} - r^2 (\dot{\theta}^2 + \sin^2 \theta \, \dot{\phi}^2 ) \right] - V(t,r,\theta,\phi)  \, ,  \nonumber
\end{eqnarray}
has  also {\it five} NSs \cite{hmq2007} for constant potential, which are {\it four KVs} given in \eqref{kv-0123} for the Schwarzschild metric and the translation symmetry ${\bf Y}_0 = \p_s$.

\subsection{Noether Symmetries of the Kerr Spacetime}

Here I will use the signature $(-,+,+,+)$ for the Kerr spacetime, in which
the line element in Boyer-Lindqust coordinates is given by
\begin{eqnarray}
& &  ds^2 = -\left( 1- \frac{2 G M r}{\Sigma c^2 } \right) c^2 dt^2 + \frac{\Sigma}{\Delta } dr^2 + \Sigma d\theta^2 - \frac{4 G M a r \sin^2\theta}{\Sigma c^2} d t d\phi   + \left[ \left( r^2 + \frac{a^2}{c^2} \right)^2 - \frac{a^2}{c^2} \Delta \sin^2\theta \right] \frac{\sin^2\theta}{\Sigma} d\phi^2 \, , \label{kerr-metric} \qquad
\end{eqnarray}
where $ \Sigma = r^2 + \frac{a^2}{c^2} \cos^2\theta$ and $ \Delta = r^2 + \frac{a^2}{c^2} - \frac{2 G M r}{c^2}$ with the mass $M$ and the spin $a = J/(M c)$ (units of length) of the gravitating source.

The geodesic Lagrangian \eqref{canonic-lagr} for the Kerr metric \eqref{kerr-metric}  is
\begin{eqnarray}
& &  \mathcal{L} = \frac{1}{2} \Big[ -\left( 1 - \frac{2 G M r}{\Sigma c^2} \right) c^2 \dot{t}^2 + \frac{ \Sigma}{\Delta} \dot{r}^2  + \Sigma \dot{\theta}^2  \nonumber \\ & &  \qquad \qquad + \left( \left( r^2 + \frac{a^2}{c^2} \right)^2 - \frac{a^2}{c^2} \Delta \sin^2\theta \right) \frac{\sin^2\theta}{\Sigma} \dot{\phi}^2  - \frac{4 G M a r \sin^2\theta}{\Sigma c^2}\, \dot{t} \, \dot{\phi} \Big] - V(t,r,\theta,\phi)  \, ,  \nonumber
\end{eqnarray}
Solving the NS equations for the geodesic Lagrangian of the Kerr metric we get {\it two isometries} and the {\it translation of the geodesic parameter} as NS generators \cite{hmq2009a}
\begin{equation}
 {\bf K}_0 = \p_t \, , \qquad {\bf K}_3 = \p_{\phi} \, , \qquad {\bf Y}_0 = \p_s \, , \label{kerr-ns}
\end{equation}
corresponding to the conservation of total energy, conservation of the angular momentum per unit mass at azimuthal direction, and the translation of the arc length, respectively.

\section{Approximate Noether Symmetries for the geodesic Lagrangians}\label{sec3}

In this section, we introduce the {\it approximate Noether symmetry} (ANS) approach of the first-order perturbed Lagrangian extending the procedure of obtaining  ANSs until the $n^{th}$-order. A {\it perturbed Lagrangian} to $n^{th}$-order can be written as
\begin{eqnarray}
& & \mathcal{L} (s, x^i, \dot{x}^i; \epsilon) = \mathcal{L}_0 (s, x^i, \dot{x}^i) + \epsilon \mathcal{L}_1 (s, x^i, \dot{x}^i) + \ldots  + \epsilon^n \mathcal{L}_n (s, x^i, \dot{x}^i) +  O(\epsilon^{n+1}) . \label{p-lagr}
\end{eqnarray}
Then an ANS generator related with the above Lagrangian is given by
\begin{equation}
 {\bf X} = {\bf X}_0 + \epsilon {\bf X}_1 + \epsilon^2 {\bf X}_2 + \ldots + \epsilon^n {\bf X}_{n}, \label{appr-noet-gen}
\end{equation}
up to the gauge function $$f(s,x^i; \epsilon) = f_0 (s,x^i) + \epsilon f_1 (s,x^i) + \epsilon^2 f_2 (s,x^i) + \ldots + \epsilon^n f_{n} (s,x^i),$$
if the ANS generator \eqref{appr-noet-gen} satisfies the {\it approximate Noether symmetry conditions}
\begin{eqnarray}
& &  {\bf X}_0^{[1]} \mathcal{L}_0 + \mathcal{L}_0 \, (D_{s}\xi_0) = D_{s} f_0 , \nonumber \\& & {\bf X}_1^{[1]} \mathcal{L}_0 + {\bf X}_0^{[1]} \mathcal{L}_1 + \mathcal{L}_0 \, (D_{s}\xi_1) + \mathcal{L}_1 \, (D_{s}\xi_0)= D_{s} f_1, \nonumber \\ & &  {\bf X}_2^{[1]} \mathcal{L}_0 + {\bf X}_1^{[1]} \mathcal{L}_1 + {\bf X}_0^{[1]} \mathcal{L}_2 +  \mathcal{L}_0 \, (D_{s}\xi_2) + \mathcal{L}_1 \, (D_{s}\xi_1) +  \mathcal{L}_2 \, (D_{s}\xi_0) = D_{s} f_2, \nonumber \\ & &  \cdots \nonumber \\& & {\bf X}_n^{[1]} \mathcal{L}_0 + {\bf X}_{n-1}^{[1]} \mathcal{L}_1 + {\bf X}_{n-2}^{[1]} \mathcal{L}_2 + \ldots + \mathcal{L}_0 \, (D_{s}\xi_n) + \mathcal{L}_1 \, (D_{s}\xi_{n-1}) + \mathcal{L}_2 \, (D_{s}\xi_{n-2}) + \ldots  = D_{s} f_n, \nonumber
\end{eqnarray}
where $n \geq 1$, and ${\bf X}_0$ is the {\it exact} NS generator,  ${\bf X}_1, {\bf X}_2, \ldots {\bf X}_n$ are the {\it first-order}, {\it second-order}, $\ldots, n^{th}$-order ANS generators, respectively, which are defined as
\begin{eqnarray}
& &  {\bf X}_{\alpha} = \xi_{\alpha} \frac{\p}{\p s} + \eta^i_{\alpha} \frac{\p}{\p x^i}, \quad (\alpha = 0,1,2,\ldots, n), \nonumber \\ & & {\bf X}_{\alpha}^{[1]} = {\bf X}_{\alpha} + \eta^i_{{\alpha}(\tau)} \frac{\partial}{\partial \dot{x}^i}, \qquad {\eta^i_{{\alpha}(\tau)}} = D_{s}\eta^i_{\alpha} - \dot{x}^i D_{s}\xi_{\alpha}  \, . \nonumber
\end{eqnarray}

The spacetime metric $g_{ij}$ can be decomposed up to $n^{th}$-order as follows:
\begin{equation}
g_{ij} = \gamma_{ij} + \epsilon h_{ij} + \epsilon^2 \sigma_{ij} + \ldots + \epsilon^n \lambda_{ij}  \, , \label{p-metric}
\end{equation}
which means by \eqref{p-lagr} and \eqref{p-metric} that the exact and perturbed geodesic Lagrangians of motion have the form
\begin{eqnarray}
& &  \mathcal{L}_0 (s, x^k, \dot{x}^k) = \frac{1}{2} \gamma_{ij} \dot{x}^i \dot{x}^j - V_0 (s, x^k) \, , \nonumber \\ & &  \mathcal{L}_1 (s, x^k, \dot{x}^k) = \frac{1}{2}  h_{ij} \dot{x}^i \dot{x}^j - V_1 (s,x^k), \nonumber \\ & &  \mathcal{L}_2 (s, x^k, \dot{x}^k) = \frac{1}{2} \sigma_{ij} \dot{x}^i \dot{x}^i - V_2 (s,x^k) \, , \nonumber \\ & & \cdots \nonumber \\ & & \mathcal{L}_n (s, x^k, \dot{x}^k) = \frac{1}{2} \lambda_{ij} \dot{x}^i \dot{x}^j - V_n (s,x^k) , \nonumber
\end{eqnarray}
where $\gamma_{ij}$, $h_{ij}$, $\sigma_{ij}$ and $\lambda_{ij}$ are the exact, the first-order, the  second-order and the $n^{th}$-order perturbed metrics, respectively. The metric $\gamma_{ij}$ should be non-degenerate (i.e., $\det ( \gamma_{ij}) \neq 0$). But the other metrics  $h_{ij}, \sigma_{ij}, \ldots, \lambda_{ij}$ can be degenerate (i.e., $ \det ( h_{ij} ) = 0, \det ( \sigma_{ij}) = 0, \ldots, \det (\lambda_{ij}) = 0$) or non-degenerate, and they represent slight deviations from flat spacetime geometry if the metric $\gamma_{ij}$ represents flat geometry.

The above perturbed Lagrangian \eqref{p-lagr} yields a $n^{th}$-order (in $\epsilon$) perturbed system of ODEs. If ${\bf X}_{\alpha}$ are the ANSs corresponding to the perturbed geodesic Lagrangians $\mathcal{L}_{\alpha} (s, x^i, \dot{x}^i)$, then
\begin{eqnarray}
& & I_0 = - \xi_0 E_{\mathcal{L}_0} + \eta^i_0 \gamma_{ij} \dot{x}^j - f_0, \nonumber \\ & & I_1 = - \xi_0 E_{\mathcal{L}_1} - \xi_1 E_{\mathcal{L}_0} + \left( \eta^i_0 h_{ij} + \eta^i_1 \gamma_{ij} \right) \dot{x}^j - f_1, \nonumber \\ & &  I_2 =  -\xi_0 E_{\mathcal{L}_2} - \xi_1 E_{\mathcal{L}_1} - \xi_2 E_{\mathcal{L}_0} + \left( \eta^i_0 \sigma_{ij}  + \eta^i_1 h_{ij} + \eta^i_2 \gamma_{ij} \right) \dot{x}^j - f_2, \label{fint-ANS}\\& &  \cdots \nonumber \\& &  I_n =  -\xi_0 E_{\mathcal{L}_n} - \xi_1 E_{\mathcal{L}_{n-1}} - \ldots - \xi_n E_{\mathcal{L}_0}  + \Big( \eta^i_0 \lambda_{ij} + \ldots + \eta^i_{n-2} \sigma_{ij} + \eta^i_{n-1} h_{ij} + \eta^i_{n} \gamma_{ij} \Big) \dot{x}^j - f_n, \nonumber
\end{eqnarray}
are the first integrals associated with ANSs ${\bf X}_{\alpha}, \,(\alpha=0,1,2,\ldots,n)$. Here the exact and the perturbed energy functionals for the perturbed Lagrangian \eqref{p-lagr} are
\begin{eqnarray}
& &  E_{\mathcal{L}_0} = \frac{1}{2} \gamma_{ij} \dot{x}^i \dot{x}^j + V_0 \, ,  \quad   E_{\mathcal{L}_1} = \frac{1}{2} h_{ij} \dot{x}^i \dot{x}^j + V_1 \, , \quad   E_{\mathcal{L}_2} = \frac{1}{2} \sigma_{ij} \dot{x}^i \dot{x}^j + V_2 \quad \, , \ldots , \quad  E_{\mathcal{L}_n} = \frac{1}{2} \lambda_{ij} \dot{x}^i \dot{x}^j + V_n  \, . \nonumber
\end{eqnarray}

It has been investigated the ANSs and conservation laws of the geodesic equations without the potential function for the Schwarzschild \cite{kmq2008} and the RN \cite{hmq2007} the spacetimes.
Constructing the geometrical set of equations corresponding to the ANS equations with an arbitrary potential function, the ANSs of the geodesic Lagrangian for the Schwarzschild, the RN and the Bardeen spacetimes have been determined by Camci \cite{camci2014a,camci2014b}.
Hussain et al. \cite{hmq2009a} have recovered all the lost conservation laws as trivial second-order approximate conservation laws of a Lagrangian for the geodesic equations by using the ANS approach in the Kerr and the charged-Kerr spacetimes. They have also discussed the problem of energy in cylindrical and plane gravitational wave spacetimes using approximate Noether symmetry method \cite{hmq2009b}. Ali and Feroze \cite{at2015} have generalized the work in Ref.\cite{hmq2009a} such that the ANS of the most general plane symmetric static spacetime are obtained.

\subsection{Approximate Symmetries of the Schwarzschild Spacetime}

First, we will look at the ANSs of geodesic Lagrangian for the Schwarzschild metric. In that context, we consider the Schwarzschild line element given in \eqref{sch-metric}.
Setting $2 G M c^{-2} = r_0\epsilon $ and using $$\left( 1 - \frac{2 G M}{c^2 r} \right)^{-1} = 1 + \frac{\epsilon\, r_0}{r} + O(\epsilon^2),$$ the first-order perturbed geodesic Lagrangian of Schwarzschild metric is given by
\begin{eqnarray}
& &  \mathcal{L} = \frac{1}{2} \left[ \dot{t}^2 - \dot{r}^2 - r^2 (\dot{\theta}^2 + \sin^2 \theta \, \dot{\phi}^2 ) \right]  - \frac{\epsilon \,r_0}{2 r} \left(  \dot{t}^2 + \dot{r}^2 \right) - V(t,r,\theta,\phi) + O(\epsilon^2), \qquad  \label{lagr-schwarz}
\end{eqnarray}
which yields
\begin{eqnarray}
& & \mathcal{L}_0 = \frac{1}{2} \left[ \dot{t}^2 - \dot{r}^2 - r^2 (\dot{\theta}^2 + \sin^2 \theta \, \dot{\phi}^2 ) \right] - V_0, \quad \mathcal{L}_1 = - \frac{r_0}{2 r} \left(  \dot{t}^2 + \dot{r}^2 \right) - V_1 , \label{lagr-schw-01}
\end{eqnarray}
where $r_0$ is a dimensional parameter (units of length), $\gamma_{ij} = {\rm diag} (1, -1, -r^2, -r^2 \sin^2 \theta)$ is called as the Minkowski metric, and $h_{ij} = {\rm diag}(-r_0/r, -r_0/r, 0,0)$. Moreover, the above Lagrangian reduces to the geodesic Lagrangian of the Minkowski metric in the limit $\epsilon = 0$.

Applying the ANS approach to these exact and perturbed metrics $\gamma_{ij}$ and $h_{ij}$ for the Schwarzschild spacetime, we find from the exact and first-order ANS equations that for the constant potential, e.g. $V(t,r,\theta,\phi) = V_0 + \epsilon V_1$ where $V_0,V_1$ are constants, we find {\it 5 exact ANSs} and {\it 17 first-order ANSs} which includes also exact ones. Here the exact ANSs are the \emph{four} KVs given in \eqref{kv-01} and \eqref{kv-23} and \emph{one} non-Killing vector field ${\bf Y}_0 = \p_s$ which gives translation in geodetic parameter $s$ and it always exists for the canonical geodesic Lagrangian \eqref{canonic-lagr}.
The remaining first-order nontrivial ANSs are
\begin{eqnarray}
& & {\bf Y}_1 = \sin \theta \cos\phi \p_r  +\frac{\cos \theta \cos \phi}{r} \p_{\theta} - \frac{\csc\theta \sin\phi}{r} \p_{\phi} \, , \nonumber \\
& & {\bf Y}_2 = \sin \theta \sin\phi \p_r  + \frac{\cos \theta \sin \phi}{r} \p_{\theta} + \frac{\csc\theta \cos\phi}{r} \p_{\phi} \, , \nonumber
\\ & & {\bf Y}_3 = \cos\theta \p_r - \frac{\sin\theta}{r} \p_{\theta} \, , \nonumber
\\ & & {\bf Y}_4 = r \sin\theta \cos\phi \p_t + t {\bf Y}_1 \, , \nonumber
\\ & & {\bf Y}_5 = r \sin\theta \sin\phi \p_t + t {\bf Y}_2 \, , \nonumber
\\ & & {\bf Y}_6 = r \cos\theta \p_t + t {\bf Y}_3 \, , \label{Y1-12}
\\ & & {\bf Y}_7 = s \p_s + \frac{1}{2} \left( t \p_t + r \p_r \right), \quad {\rm with \,\,} f_1 = - V_0 s \, , \nonumber
\\ & &  {\bf Y}_8 = s {\bf K}_0 \, , \quad \,\, {\rm with \,\, } f_1 = t \, , \nonumber
\\ & & {\bf Y}_9 = s {\bf Y}_1 \, , \quad \,\, {\rm with \,\, } f_1 = -r \sin\theta \cos\phi \, , \nonumber
\\ & & {\bf Y}_{10} = s {\bf Y}_2 \, , \quad {\rm with \,\, } f_1 = -r \sin\theta \sin\phi \, , \nonumber
\\ & & {\bf Y}_{11} = s {\bf Y}_3 \, , \quad {\rm with \,\, } f_1 = -r \cos\theta \, , \nonumber
\\ & & {\bf Y}_{12} = s \left( s \p_s + t \p_t + r \p_r \right) \, , \,\, {\rm with \,\, } f_1 = \frac{1}{2} \left( t^2 -r^2 - 2 V_0 s^2 \right) \, . \nonumber
\end{eqnarray}
For the Schwarzschild spacetime considered as a first perturbation of the Minkowski metric,
\emph{three} nontrivial ANS generators ${\bf Y}_1, {\bf Y}_2, {\bf Y}_3$ provide the {\it conservation of linear momentum}, and \emph{three} nontrivial ANS generators ${\bf Y}_4, {\bf Y}_5, {\bf Y}_6$  give the conservation of spin angular momentum due to Lorentz invariance.

\bigskip

The first integrals associated with the {\it 5 exact} ANSs are
\begin{eqnarray}
I_0^1 = - E_{\mathcal{L}_0} , \, I_0^2 = \dot{t}, \, I_0^3 = - r^2 \sin^2\theta \, \dot{\phi} , \,    I_0^4 = -r^2 \left[ \cos\phi \, \dot{\theta} - \sin\theta \cos\theta \sin\phi \, \dot{\phi} \right], \,   I_0^5 = -r^2 \left[ \sin\phi \, \dot{\theta} + \sin\theta \cos\theta \cos\phi \, \dot{\phi} \right], \quad & & \label{exact-fint}
\end{eqnarray}
where the {\it exact energy functional} $E_{\mathcal{L}_0}$ is
\begin{equation}
  E_{\mathcal{L}_0} = \frac{1}{2} \left[ \dot{t}^2 - \dot{r}^2 - r^2 (\dot{\theta}^2 + \sin^2\theta \, \dot{\phi}^2 ) \right] + V_0. \label{sch-el0}
\end{equation}
Then the first integrals associated with the {\it 17 first-order} ANSs are given by
\begin{eqnarray}
& &  I_1^1 = - (E_{\mathcal{L}_0} + E_{\mathcal{L}_1} ), \quad I_1^2 = \left( 1 - \frac{r_0}{r} \right) \dot{t}, \quad I_1^3 = - r^2 \sin^2\theta \, \dot{\phi} ,  \nonumber \\ & &  I_1^4 = -r^2 \left[ \cos\phi \, \dot{\theta} - \sin\theta \cos\theta \sin\phi \, \dot{\phi} \right], \,\,  I_1^5 = -r^2 \left[ \sin\phi \, \dot{\theta} + \sin\theta \cos\theta \cos\phi \, \dot{\phi} \right],  \nonumber \\ & &  I_1^6 = -\sin\theta \cos\phi \, \dot{r} - r \cos\theta \cos\phi \, \dot{\theta} + r \sin\theta \sin\phi \, \dot{\phi}, \nonumber \\ & &  I_1^7 = -\sin\theta \sin\phi \, \dot{r} - r \cos\theta \sin\phi \, \dot{\theta} - r \sin\theta \cos\phi \, \dot{\phi}, \nonumber \\ & &  I_1^8 = - \cos\theta \, \dot{r} + r \sin\theta \, \dot{\theta}, \qquad I_1^9 = r \sin\theta \cos\phi \, \dot{t} + I_1^6 t, \label{first-order-fint} \\ & &  I_1^{10} = r \sin\theta \sin\phi \, \dot{t} + I_1^7 t, \quad \,\,\, I_1^{11} = r \cos\theta  \, \dot{t} + I_1^8 t, \nonumber \\ & &  I_1^{12} = s \, \dot{t} - t  \, ,  \quad I_1^{13} = - ( E_{\mathcal{L}_0} - V_0) s + \frac{1}{2} ( t \, \dot{t} - r \, \dot{r})  \, ,  \nonumber \\ & &  I_1^{14} = - ( E_{\mathcal{L}_0} - V_0) s^2 + s ( t \, \dot{t} - r \, \dot{r}) - \frac{1}{2} (t^2 -r^2) \, ,  \nonumber \\ & &  I_1^{15} = I_1^6 s + r \sin\theta \cos\phi \, , \quad I_1^{16} = I_1^7 s + r \sin\theta \sin\phi \, , \quad I_1^{17} = I_1^8 s + r \cos\theta,  \nonumber
\end{eqnarray}
where the {\it first-order energy functional} $E_{\mathcal{L}_1}$ is
\begin{equation}
  E_{\mathcal{L}_1} = -\frac{r_0}{2 r} \left( \dot{t}^2 + \dot{r}^2 \right) + V_1. \label{sch-el1}
\end{equation}
Defining the Noether constants as $I_1^2 = - E$, $I_1^3 = L_z$, $I_1^4 = a_1$, $I_1^5 = a_2$, $I_1^6 = a_3$, $I_1^7 = a_4$, $I_1^8 = a_5$, $I_1^9 = a_6$, $I_1^{10} = a_7$, $I_1^{11} = a_8$, $I_1^1 = b_1$, $I_1^{12} = b_2$, $I_1^{13} = b_3$, $I_1^{14} = b_4$, $I_1^{15} = b_5$, $I_1^{16} = b_6$, $I_1^{17} = b_7$, and using the first integrals \eqref{first-order-fint}, it follows that
\begin{eqnarray}
& &  E_{\mathcal{L}_0} + E_{\mathcal{L}_1} = - b_1,  \quad E = -t_0 \left( 1 - \frac{r_0}{r} \right), \quad L_z = - r^2 \sin^2\theta \, \dot{\phi} , \label{fint-sch1} \\ & &  t (s) = t_0 \, s - b_2 , \qquad t_0 = \frac{1}{b_2} ( a_3 b_5 + a_4 b_6 + a_5 b_7 - 2 b_3), \label{fint-sch2} \\ & &  r \sin\theta \cos\phi = -a_3 s + b_5, \,\, r \sin\theta \sin\phi = - a_4 s + b_6, \,\, r \cos\theta = - a_5 s + b_7 , \qquad \label{fint-sch3} \\ & &  a_1 = a_3 b_7 - a_5 b_5, \quad a_2 = a_4 b_7 - a_5 b_6, \quad b_4 = \frac{1}{2} \left( b_5^2 + b_6^2 + b_7^2 - b_2^2 \right), \label{fint-sch4}  \\ & &  a_6 = t_0 b_5 - a_3 b_2, \quad a_7 = t_0 b_6 - a_4 b_2, \quad a_8 = t_0 b_7 - a_5 b_2 ,  \label{fint-sch5}
\end{eqnarray}
where $t_0$ is a constant of integration and $b_2 \neq 0$. Thus, Eq. \eqref{fint-sch3} yields
\begin{eqnarray}
& &  r(s) = \sqrt{ (a_3 s - b_5)^2 + (a_4 s - b_6)^2 + (a_5 s - b_7)^2 }, \label{sch-sol2} \\ & &  \theta (s) = \tan^{-1} \left( \frac{\sqrt{(a_3 s - b_5)^2 + (a_4 s - b_6)^2}}{-a_5 s + b_7} \right),  \,\, \phi (s) = \tan^{-1} \left( \frac{a_4 s - b_6}{a_3 s - b_5} \right). \qquad \label{sch-sol3}
\end{eqnarray}
After considering Eqs. \eqref{sch-el0} and \eqref{sch-el1}, we have found the exact and the first-order energy functionals as
\begin{eqnarray}
& &  E_{\mathcal{L}_0} = \frac{1}{2} \left( t_0^2 - a_3^2 - a_4^2 - a_5^2 \right) + V_0 \, , \label{schw-EL0} \\ & &  E_{\mathcal{L}_1} = -\frac{r_0}{2 r} \left[  t_0^2 + \frac{1}{r^2} \left[ \left(a_3^2 + a_4^2 + a_5^2 \right) s - (a_3 b_5 + a_4 b_6 + a_5 b_7) \right]^2 \right] + V_1 \, . \qquad \label{schw-EL1}
\end{eqnarray}
Further, using Eqs. \eqref{sch-sol2} and \eqref{sch-sol3}, it follows from Eq.\eqref{fint-sch1} that the component of angular momentum $L_z$ becomes a constant such as $L_z = a_4 b_5 - a_3 b_6$. We point out the fact that the exact energy functional $E_{\mathcal{L}_0}$ given in \eqref{schw-EL0} is already a constant. It is seen from Eq. \eqref{schw-EL1} if $a_3^2 + a_4^2 + a_5^2 = 0$ and $a_3 b_5 + a_4 b_6 + a_5 b_7 = 0$, i.e., this means $r = \sqrt{ b_5^2 + b_6^2 + b_7^2 } =const.$, then the first-order energy functional $E_{\mathcal{L}_1}$ becomes constant, i.e. $E_{\mathcal{L}_1}= - \frac{r_0 t_0^2}{2 \, r} + V_1$.

\subsection{Approximate Symmetries of the Reissner-Nordstr\"{o}m Spacetime}

Setting $2 G M c^{-2} = r_0 \, \epsilon $ and $G Q^2 c^{-4} = q \epsilon^2$ at the RN spacetime \eqref{rn-metric}, the RN metric coefficients become
\begin{eqnarray}
& & g_{tt} = 1 - \frac{\epsilon \, r_0}{r} + \frac{k \epsilon^2}{r^2} + O(\epsilon^3)  \quad {\rm and} \quad  g_{rr} = -\left[ 1 + \frac{\epsilon \, r_0}{r} + (1-q) \frac{\epsilon^2}{r^2}\right] + O(\epsilon^3). \nonumber
\end{eqnarray}
Therefore, the second-order perturbed geodesic Lagrangian of the RN metric has the form:
\begin{eqnarray}
& &  \mathcal{L} = \frac{1}{2} \left[ \dot{t}^2 - \dot{r}^2 - r^2 (\dot{\theta}^2 + \sin^2 \theta \, \dot{\phi}^2 ) \right]  - \frac{\epsilon \, r_0}{2 r} \left(  \dot{t}^2 + \dot{r}^2 \right)  + \frac{\epsilon^2}{2 r^2} \left[ q \dot{t}^2 + (q-1) \dot{r}^2 \right] - V(t,r,\theta,\phi) + O(\epsilon^3), \label{RN-lagr}
\end{eqnarray}
which yields the same $\mathcal{L}_0$ and $\mathcal{L}_1$ given for the Schwarzschild spacetime, and
\begin{eqnarray}
& &  \mathcal{L}_2 =  \frac{1}{2 r^2} \left[ q \dot{t}^2 + (q-1) \dot{r}^2 \right] - V_2, \label{RN-lagr2}
\end{eqnarray}
where $\gamma_{ij} = {\rm diag} (1, -1, -r^2, -r^2 \sin^2 \theta)$, $h_{ij} = {\rm diag} \left( -r_0 /r, -r_0 /r, 0,0 \right) $ and $\sigma_{ij} = {\rm diag} \left( q / r^2, (q-1)/r^2,0,0 \right) $.
Retaining only the first order terms in the above Lagrangian and neglecting $O(\epsilon^2)$, it reduces to the first-order perturbed geodesic Lagrangian for the Schwarzschild metric.

It is seen from the solutions of ANS equations of the RN metric that the exact ANSs of the Schwarzschild metric are retained, i.e., there exist {\it 5 exact} ANS generators, which are ${\bf Y}_0, {\bf K}_0,{\bf K}_1,{\bf K}_2$ and ${\bf K}_3$. There exist also {\it 5 first-order} ANSs for the RN metric as for the exact ones. The {\it lost} symmetries for the RN metric appear in the second-order ANS generators which are solutions of ANS conditions with the constant potential, and the number of the second-order nontrivial ANS generators is seventeen, which are the same with ${\bf K}_0,\ldots,{\bf K}_3, {\bf Y}_0,\ldots,{\bf Y}_{12}$ given the symmetry generators for the first-order perturbed Schwarzschild metric. The first integrals of the second-order ANSs for the RN metric have the same form given by the Schwarzschild metric. In summary, the solutions for these first integrals are as follows:
\begin{eqnarray}
& &  t(s) = t_0\, s - b_2 \, , \quad
r(s) = \sqrt{ (a_3 s - b_5)^2 + (a_4 s - b_6)^2 + (a_5 s - b_7)^2 }, \label{rn-r} \\ & & \theta (s) = \tan^{-1} \left( \frac{\sqrt{(a_3 s - b_5)^2 + (a_4 s - b_6)^2}}{-a_5 s + b_7} \right), \,\,  \phi (s) = \tan^{-1} \left( \frac{a_4 s - b_6}{a_3 s - b_5} \right),
\end{eqnarray}
together with the constraints depending on the Noether constants,
\begin{eqnarray}
& &  E_{\mathcal{L}_0} + E_{\mathcal{L}_1} + E_{\mathcal{L}_2} = - b_1 \, , \,\, E = -t_0 \left( 1 - \frac{r_0}{r} + \frac{q}{r^2} \right) \, , \,\, L_z = a_4 b_5 - a_3 b_6 \, , \qquad \label{fint-rn1} \\ & & t_0 = \frac{1}{b_2} ( a_3 b_5 + a_4 b_6 + a_5 b_7 - 2 b_3) \, ,   \label{fint-rn2} \\ & &  a_1 = a_3 b_7 - a_5 b_5 \, , \qquad a_2 = a_4 b_7 - a_5 b_6 \, , \qquad b_4 = \frac{1}{2} \left( b_5^2 + b_6^2 + b_7^2 - b_2^2 \right),  \label{fint-rn3}  \\ & &  a_6 = t_0 b_5 - a_3 b_2 \, , \qquad a_7 = t_0 b_6 - a_4 b_2 \, , \qquad a_8 = t_0 b_7 - a_5 b_2 \, .  \label{fint-rn4}
\end{eqnarray}
Here the exact and first-order energy functionals are the same with \eqref{schw-EL0} and \eqref{schw-EL1}, and the second-order energy functional $E_{\mathcal{L}_2}$ reads
\begin{equation}
  E_{\mathcal{L}_2} = \frac{1}{2 r^2} \left[ q\, t_0^2 + \frac{(q-1)}{r^2} \left[ \left(a_3^2 + a_4^2 + a_5^2 \right) s - (a_3 b_5 + a_4 b_6 + a_5 b_7) \right]^2 \right] + V_2 .  \label{rn-EL2}
\end{equation}
We point out again that for the RN metric the component of angular momentum $L_z$ and the exact energy functional $E_{\mathcal{L}_0}$ are constants. But, the energy $E$ and the energy functionals $E_{\mathcal{L}_1}, E_{\mathcal{L}_2}$ are explicitly depend on the arc length $s$. Further, using the first and second relations given in Eq. \eqref{fint-rn1}, we have found the energy $E$ as
\begin{eqnarray}
& &  E = \frac{2}{t_0} \left[ b_1 - \frac{1}{2}( a_3^2 + a_4^2 + a_5^2) + V_0 + V_1 + V_2 \right]  \nonumber \\ & & \qquad \quad +  \frac{1}{t_0 r(s)^3} \left( -r_0 + \frac{q-1}{r(s)} \right) \left[ \left(a_3^2 + a_4^2 + a_5^2 \right) s - (a_3 b_5 + a_4 b_6 + a_5 b_7) \right]^2   \, ,
\end{eqnarray}
where $r(s)$ is of the form \eqref{rn-r}.

\subsection{Approximate Symmetries of the Kerr Spacetime}

In Boyer-Lindqust coordinates the Kerr spacetime is given in \eqref{kerr-metric}, where
\begin{eqnarray}
& &  \Sigma = r^2 + \frac{a^2}{c^2} \cos^2\theta, \qquad \Delta = r^2 + \frac{a^2}{c^2} - \frac{2 G M r}{c^2} \, ,
\end{eqnarray}
with the mass $M$ and the spin parameter $a$. Introducing the small parameter $\epsilon$ as
\begin{equation}
   \frac{2 G M}{c^2} =  r_0 \, \epsilon \, , \qquad \frac{a}{c} =  k \epsilon \, ,
\end{equation}
and retaining third power of $\epsilon$ and neglecting higher powers, the metric coefficients
in the Kerr spacetime become
\begin{eqnarray}
& &  g_{tt} = - 1 + \frac{\epsilon \, r_0}{r} - \frac{r_0 k^2 }{r^3} \epsilon^3 \cos^2\theta  + O(\epsilon^4) \, , \\ & &  g_{rr} = 1 + \frac{\epsilon \, r_0}{r} + \left( r_0^2 - k^2 \sin^2\theta \right) \frac{\epsilon^2}{r^2} +  \left( r_0^2 - 2 k^2 + k^2 \cos^2\theta \right) \frac{\epsilon^3}{r^3} + O(\epsilon^4), \qquad \\ & &  g_{\theta \theta} = r^2 + \epsilon^2 k^2 \cos^2\theta , \quad  g_{t \phi} = -\frac{k r_0}{r} \epsilon^2 \sin^2\theta  + O(\epsilon^4) \, , \\ & &  g_{\phi \phi} = \sin^2\theta \left( r^2 + k^2 \epsilon^2 + \frac{r_0 k^2}{r} \epsilon^3 \sin^2\theta  \right) + O(\epsilon^4) \, .
\end{eqnarray}
The third-order perturbed geodesic Lagrangian for the Kerr spacetime is given by
\begin{equation}
  \mathcal{L} = \mathcal{L}_0 + \epsilon \mathcal{L}_1 + \epsilon^2 \mathcal{L}_2 + \epsilon^3 \mathcal{L}_3 + O(\epsilon^4) \, ,
\end{equation}
where the exact, first-order, second-order and third-order geodesic Lagrangians are as follows:
\begin{eqnarray}
& & \mathcal{L}_0 = \frac{1}{2} \left[ -\dot{t}^2 + \dot{r}^2 + r^2 (\dot{\theta}^2 + \sin^2 \theta \, \dot{\phi}^2 ) \right] - V_0, \quad \mathcal{L}_1 = \frac{r_0}{2 r} \left(  \dot{t}^2 + \dot{r}^2 \right) - V_1 , \nonumber \\ & &   \mathcal{L}_2 = \frac{1}{2} \left[ \frac{( r_0^2 - k^2\sin^2\theta)}{r^2} \dot{r}^2 + k^2 \cos^2\theta \, \dot{\theta}^2 + \sin^2\theta \left( k^2 \, \dot{\phi}^2  - \frac{2 k r_0}{r} \, \dot{t} \, \dot{\phi} \right) \right] - V_2, \nonumber \\ & &  \mathcal{L}_3 = \frac{1}{2} \left[ - \frac{r_0 k^2}{r^3} \cos^2\theta \, \dot{t}^2  + \frac{( r_0^2 - 2 k^2 + k^2\cos^2\theta)}{r^3} \dot{r}^2 + \frac{r_0 k^2}{r} \sin^4\theta \, \dot{\phi}^2  \right] - V_3. \nonumber
\end{eqnarray}

There are only {\it three} ANS generators for the exact geodesic Lagrangian of the geodesic equations of Kerr spacetime such as
\begin{equation}
{\bf K}_0 = \p_t \, , \qquad {\bf K}_3 = \p_{\phi} \, , \qquad {\bf Y}_0 = \p_s \, . \label{ans-kerr-1}
\end{equation}
This is also pointed out by the Ref. \cite{hmq2009a},
where they have proceeded the ANS to the second-order ANS of the geodesic Lagrangian for the Kerr spacetime. After proceeding the ANS to the third-order ANS, the solution of first-order ANS equations for constant potential yields {\it three} ANSs as in \eqref{ans-kerr-1}, and the {\it two additional} ANS generators arise for the {\it second-order} approximation as the following
\begin{eqnarray}
& & {\bf K}_1 = \cos \phi \p_{\theta} - \cot \theta \sin \phi \p_{\phi} \, , \qquad {\bf K}_2 = \sin \phi \p_{\theta} + \cot\theta \cos\phi \p_{\phi} \, .  \label{ans-kerr-k23}
\end{eqnarray}
In addition to the symmetries ${\bf K}_0, {\bf K}_1, {\bf K}_2, {\bf K}_3$ and ${\bf Y}_0$, the lost symmetries of the Kerr spacetime are appeared as the solution of the {\it third-order} ANS equations such that the symmetries ${\bf Y}_1, \ldots , {\bf Y}_7$ are the same as given in \eqref{Y1-12}, and the remaining ones are
\begin{eqnarray}
& &   {\bf Y}_8 = s {\bf K}_0 , \quad \,\, {\rm with \,\, } f_3 = -t, \nonumber \\& &   {\bf Y}_9 = s {\bf Y}_1, \quad \,\, {\rm with \,\, } f_3 = r \sin\theta \cos\phi, \nonumber \\& &   {\bf Y}_{10} = s {\bf Y}_2, \quad {\rm with \,\, } f_3 = r \sin\theta \sin\phi, \label{kerr-Y1-12} \\& &   {\bf Y}_{11} = s {\bf Y}_3, \quad {\rm with \,\, } f_3 = r \cos\theta, \nonumber \\& &  {\bf Y}_{12} = s \left( s \p_s + t \p_t + r \p_r \right), \,\, {\rm with \,\, } f_3 = \frac{1}{2} \left( r^2 - t^2 - 2 V_0 s^2 \right). \nonumber
\end{eqnarray}
Hence, the number of third-order ANSs for the geodesic Lagrangian of the Kerr spacetime is {\it seventeen}. The first integrals associated with the {\it three exact and first-order} ANSs are
\begin{equation}
   I_0^1 = - E_{\mathcal{L}_0} , \quad I_0^2 = - \dot{t}, \quad I_0^3 = r^2 \sin^2\theta \, \dot{\phi} \, ,  \label{exact-fint-kerr}
\end{equation}
and
\begin{equation}
   I_1^1 = - (E_{\mathcal{L}_0} + E_{\mathcal{L}_1} ), \quad I_1^2 = - \left( 1 - \frac{r_0}{r} \right) \dot{t}, \quad I_1^3 = r^2 \sin^2\theta \, \dot{\phi} \, ,  \label{first-fint-kerr}
\end{equation}
where the exact and first-order energy functionals are, respectively,
\begin{eqnarray}
& &    E_{\mathcal{L}_0} = \frac{1}{2} \left[ -\dot{t}^2 + \dot{r}^2 + r^2 (\dot{\theta}^2 + \sin^2\theta \, \dot{\phi}^2 ) \right] + V_0  \qquad {\rm and} \qquad
  E_{\mathcal{L}_1} = \frac{r_0}{2 r} \left(  \dot{t}^2 + \dot{r}^2 \right)  + V_1 \, . \label{first-kerr-EL0-EL1}
\end{eqnarray}
The conservation laws for the second-order ANSs of Kerr spacetime are found as:
\begin{eqnarray}
& &   I_2^1 = - \left( E_{\mathcal{L}_0} + E_{\mathcal{L}_1} + E_{\mathcal{L}_2} \right) \, , \qquad I_2^2 = \left( \frac{r_0}{r} - 1 \right) \dot{t} - \frac{k r_0}{r} \sin^2\theta \, \dot{\phi} \, , \qquad  I_2^3 =  - \frac{k r_0}{r} \dot{t} + (r^2 + k^2) \sin^2\theta \, \dot{\phi} \, ,  \nonumber \\ & & I_2^4 = r^2 \left[ \cos\phi \, \dot{\theta} - \sin\theta \cos\theta \sin\phi \, \dot{\phi} \right]  \, , \qquad  I_2^5 =  r^2 \left[ \sin\phi \, \dot{\theta} + \sin\theta \cos\theta \cos\phi \, \dot{\phi} \right] ,
\end{eqnarray}
where the second-order energy functional has the form
\begin{equation}
   E_{\mathcal{L}_2} = \frac{1}{2} \left[ \frac{( r_0^2 - k^2\sin^2\theta)}{r^2} \dot{r}^2 + k^2 \cos^2\theta \, \dot{\theta}^2 + \sin^2\theta \left( k^2 \, \dot{\phi}^2  - \frac{2 k r_0}{r} \, \dot{t} \, \dot{\phi} \right) \right] + V_2 \, . \label{second-kerr-EL2}
\end{equation}
Further, the first integrals associated with the {\it 17 third-order} ANSs are
\begin{eqnarray}
& &  I_3^1 = - (E_{\mathcal{L}_0} + E_{\mathcal{L}_1} + E_{\mathcal{L}_2} + E_{\mathcal{L}_3}) \, ,  \qquad  I_3^2 = \left(- 1 + \frac{r_0}{r} - \frac{r_0 k^2}{r^3} \cos^2\theta \right) \dot{t} - \frac{r_0 k}{r} \sin^2\theta \, \dot{\phi} \,  , \nonumber \\ & &   I_3^3 =  -\frac{r_0 k}{r} \sin^2\theta \, \dot{t} + \left( r^2 + k^2 + \frac{r_0 k^2}{r} \sin^2\theta \right) \sin^2\theta \, \dot{\phi} ,  \nonumber \\ & & I_3^4 = r^2 \left[ \cos\phi \, \dot{\theta} - \sin\theta \cos\theta \sin\phi \, \dot{\phi} \right] \, , \qquad  I_3^5 = r^2 \left[ \sin\phi \, \dot{\theta} + \sin\theta \cos\theta \cos\phi \, \dot{\phi} \right],  \nonumber \\ & &  I_3^6 = \sin\theta \cos\phi \, \dot{r} + r \cos\theta \cos\phi \, \dot{\theta} - r \sin\theta \sin\phi \, \dot{\phi}, \quad  I_3^7 = \sin\theta \sin\phi \, \dot{r} + r \cos\theta \sin\phi \, \dot{\theta} + r \sin\theta \cos\phi \, \dot{\phi}, \label{third-order-fint-kerr} \\ & & I_3^8 =  \cos\theta \, \dot{r} - r \sin\theta \, \dot{\theta}, \qquad I_3^9 = -r \sin\theta \cos\phi \, \dot{t} + I_3^6 t,  \,\, I_3^{10} = -r \sin\theta \sin\phi \, \dot{t} + I_3^7 t, \quad I_3^{11} = -r \cos\theta  \, \dot{t} + I_3^8 t,  \nonumber \\ & &   I_3^{12} = -s \, \dot{t} + t  \, ,  \quad I_3^{13} = - ( E_{\mathcal{L}_0} - V_0) s + \frac{1}{2} ( - t \, \dot{t} + r \, \dot{r})  \, ,  \quad I_3^{14} = - ( E_{\mathcal{L}_0} - V_0) s^2 + s ( - t \, \dot{t} + r \, \dot{r}) - \frac{1}{2} (r^2 -t^2) \, , \nonumber \\ & &   I_3^{15} = I_3^6 s - r \sin\theta \cos\phi, \qquad I_3^{16} = I_3^7 s - r \sin\theta \sin\phi, \qquad I_3^{17} = I_3^8 s - r \cos\theta,  \nonumber
\end{eqnarray}
where the third-order energy functional $E_{\mathcal{L}_3}$ is
\begin{equation}
   E_{\mathcal{L}_3} = \frac{1}{2} \left[ - \frac{r_0 k^2}{r^3} \cos^2\theta \, \dot{t}^2  + \frac{( r_0^2 - 2 k^2 + k^2\cos^2\theta)}{r^3} \dot{r}^2 + \frac{r_0 k^2}{r} \sin^4\theta \, \dot{\phi}^2  \right] + V_3 \, . \label{third-kerr-EL3}
\end{equation}
Defining $ I_3^2 = - E$, $I_3^3 = L_z$, $I_3^4 = a_1$, $I_3^5 = a_2$, $I_3^6 = a_3$, $I_3^7 = a_4$,  $I_3^8 = a_5$, $I_3^9 = a_6$, $I_3^{10} = a_7$, $I_3^{11} = a_8$, $I_3^1 = b_1$, $I_3^{12} = b_2$, $I_3^{13} = b_3$, $I_3^{14} = b_4$,  $I_3^{15} = b_5$, $I_3^{16} = b_6$ and $I_3^{17} = b_7$, one can find from the first integrals given by \eqref{third-order-fint-kerr} that
\begin{eqnarray}
& &  t(s) = t_0\, s + b_2 \, , \qquad
r(s) = \sqrt{ (a_3 s - b_5)^2 + (a_4 s - b_6)^2 + (a_5 s - b_7)^2 } \, , \label{t-r-kerr}  \\ & & \theta (s) = \tan^{-1} \left( \frac{\sqrt{(a_3 s - b_5)^2 + (a_4 s - b_6)^2}}{a_5 s  - b_7} \right) \, , \qquad  \phi (s) = \tan^{-1} \left( \frac{a_4 s - b_6}{a_3 s - b_5} \right) \, , \label{theta-phi-kerr}
\end{eqnarray}
and
\begin{eqnarray}
& &  E = t_0 \left[ 1 - \frac{m}{r} + \frac{m k^2}{r^5} \left( a_5 s - b_7 \right)^2 \right]  +  \frac{k r_0 L_0}{r^3} \,  , \label{fint-kerr-1} \\ & & L_z =  \frac{k r_0}{r^3} \left( \frac{k L_0}{r^2} - t_0 \right) \left[ r^2 - \left(a_5 s - b_7\right)^2 \right]  + L_0 \left( 1 + \frac{k^2}{r^2} \right) \, , \label{fint-kerr-2} \\ & &  E_{\mathcal{L}_0} + E_{\mathcal{L}_1} + E_{\mathcal{L}_2} + E_{\mathcal{L}_3}  = - b_1 \, , \label{fint-kerr-3} \\ & &  t_0 = -\frac{1}{b_2} ( a_3 b_5 + a_4 b_6 + a_5 b_7 + 2 b_3) \, ,  \,\, b_2 \neq 0, \qquad b_4 = \frac{1}{2} \left( b_2^2 - b_5^2 - b_6^2 - b_7^2 \right),  \label{fint-kerr-4} \\ & &  a_1 = a_5 b_5 - a_3 b_7 \, , \quad a_2 = a_5 b_6 - a_4 b_7 \, , \quad  a_6 = t_0 b_5 + a_3 b_2 \, , \quad a_7 = t_0 b_6 + a_4 b_2 \, , \quad a_8 = t_0 b_7 + a_5 b_2 ,   \label{fint-kerr-5}
\end{eqnarray}
where $t_0$ is a constant of integration, $b_2 \neq 0$ and $L_0 \equiv a_3 b_6 - a_4 b_5 $.
Using \eqref{t-r-kerr}-\eqref{theta-phi-kerr} in Eqs. \eqref{first-kerr-EL0-EL1},  \eqref{second-kerr-EL2} and \eqref{third-kerr-EL3}, the exact, first-order, second-order and third order energy functionals for the Kerr spacetime take the following forms:
\begin{eqnarray}
& &  E_{\mathcal{L}_0} = \frac{1}{2} \left( -t_0^2 + a_3^2 + a_4^2 + a_5^2 \right) + V_0 \, , \qquad E_{\mathcal{L}_1} = \frac{r_0}{2 r} \left(  t_0^2 + \dot{r}^2  \right) + V_1 , \qquad \label{kerr-EL01} \\ & &   E_{\mathcal{L}_2} = \frac{1}{2 r^2} \left[ r_0^2 - k^2 + \frac{k^2}{r^2} (a_5 s - b_7)^2 \right] \dot{r}^2  + \frac{k^2}{2 r^2} (a_5 s - b_7)^2 \, \dot{\theta}^2  + \frac{k^2 L_0^2}{2 r^2 \left[ r^2 - \left(a_5 s - b_7\right)^2 \right] } -  \frac{k r_0 t_0 L_0}{ r^3 }  + V_2  \, , \label{kerr-EL2} \\ & &  E_{\mathcal{L}_3} =  \frac{1}{2 r^5} \Big\{ r_0 k^2 \left[ L_0^2 - t_0^2 (a_5 s - b_7)^2 \right]  + \left[ (r_0^2- 2 k^2) r^2 + k^2 (a_5 s - b_7)^2  \right] \, \dot{r}^2   \Big\}  + V_3 \, , \quad \label{kerr-EL3}
\end{eqnarray}
where $\dot{r} = \frac{1}{r} \left[ (a_3^2 + a_4^2 + a_5^2 ) s - (a_3 b_5 + a_4 b_6 + a_5 b_7) \right],$  and $$\dot{\theta}^2 = \frac{1}{r^4 \left[ r^2 - (a_5 s - b_7)^2 \right]} \left( \left[ a_5 L_1 - b_7 (a_3^2 + a_4^2) \right] s - a_5 (b_5^2 + b_6^2 ) + b_7 L_1 \right)^2 \, ,$$
with $ L_1 = a_3 b_5 + a_4 b_6$.
The {\it photon orbits} staying at the extrema, i.e., the circular equatorial orbits or the spherical photon orbits, imply $\dot{r} = 0 $ and $\ddot{r} = 0$, from which one can find the following constraint equations:
\begin{equation}
   a_3^2 + a_4^2 + a_5^2 = 0 \, , \qquad a_3 b_5 + a_4 b_6 + a_5 b_7 = 0 \, .
\end{equation}
The latter equations yield $r = \sqrt{ b_5^2 + b_6^2 + b_7^2 } = {\rm const.}$
Then the approximate energy functionals for the photon orbits read
\begin{eqnarray}
& &  E_{\mathcal{L}_0} = -\frac{t_0^2}{2} + V_0, \qquad E_{\mathcal{L}_1} = \frac{r_0 t_0^2}{2 r} + V_1 \, , \\ & &   E_{\mathcal{L}_2} = \frac{k^2}{2 r^2} (a_5 s - b_7)^2 \, \dot{\theta}^2 + \frac{k^2 L_0^2}{2 r^2 \left[ r^2 - \left(a_5 s - b_7\right)^2 \right] } -  \frac{k r_0 t_0 L_0}{ r^3 } + V_2 \, , \\ & & E_{\mathcal{L}_3} =  \frac{r_0 k^2}{2 r^5} \left[ L_0^2 - t_0^2 (a_5 s - b_7)^2 \right] + V_3 \, ,
\end{eqnarray}
where $\dot{\theta} = a_5 ( b_5^2 + b_6^2 + b_7^2)/( r^2 \left[ r^2 - (a_5 s - b_7)^2 \right] )$. It is seen from the above relations that if $a_5 = 0$, then $\theta$ is a constant and so all energy functionals become constants.

\section{Summary and Conclusion}\label{conc:sec4}

In this study, for the G\"{o}del-type, Schwarzschild, Reissner-Nordstr\"{o}m and Kerr spacetimes, we reviewed the Noether symmetries of the corresponding canonical geodesic Lagrangians. To get the approximate Lagrangian in the background of some of those spacetimes, we set up a perturbed geodesic Lagrangian in terms of metric coefficients to use it in the ANS approach. Thus we considered the latter perturbed Lagrangians and used it to calculate and classify ANS generators by considering the ANS conditions for the Schwarzschild, Reissner-Nordstr\"{o}m and Kerr spacetimes.
In the previous section, the ANSs of the Schwarzschild, Reissner-Nordstr\"{o}m  and Kerr spacetimes have been calculated, and used to integrate the geodesic equations of motion by means of the first integrals that are due to the existence of ANS generators including the KVs. The analytical solutions for the perturbed geodesic equations of these black hole spacetimes have been derived by the aid of the first integrals associated with ANSs in complete detail.

Furthermore we note that the geodesic Lagrangian involves the potential function $V(x^k)$ which is an unknown quantity. Using the Noether symmetry approach or the approximate Noether symmetry approach for the geodesic Lagrangian, the form of the unknown potential function may be determined.

\section*{Acknowledgments}

The author would like to thank the organizers for the successful meeting ``International Conference on Gravitation and Cosmology (PUICGC)", The University of Punjab, Department of Mathematics, Lahore-Pakistan held in November 22-25, 2021. I dedicate this study to Prof. Dr. Ghulam Shabbir.



\end{document}